\documentclass{aa}

\usepackage{graphicx}
\usepackage{txfonts}
\usepackage{natbib}
\usepackage{color}
\bibpunct{(}{)}{;}{a}{}{,}

\begin{document}

  \title{The bispectrum covariance beyond Gaussianity}
  \subtitle{A log-normal approach}
  \author{Sandra\ Martin, Peter\ Schneider, \and Patrick\ Simon}
  \offprints{S. Martin,\\ \email{martin@astro.uni-bonn.de}}
  \institute{Argelander-Institut f\"ur Astronomie (AIfA), Universit\"at Bonn, Auf dem H\"ugel 71, 53121 Bonn, Germany}
  \date{\today}
 
  \authorrunning{Martin et al.}
  \titlerunning{The bispectrum covariance beyond Gaussianity}

  \abstract
  {To investigate and specify the statistical properties of cosmological fields with particular attention to possible non-Gaussian features, accurate formulae for the bispectrum and
   the bispectrum covariance are required. The bispectrum is the lowest-order statistic providing an estimate for non-Gaussianities of a distribution, and the bispectrum covariance depicts the errors of the bispectrum measurement and their correlation on different scales. Currently, there do exist fitting formulae for the bispectrum and an analytical expression for the bispectrum covariance, but the former is not very accurate and the latter contains several intricate terms and only one of them can be readily evaluated from the power spectrum of the studied field. Neglecting all higher-order terms results in the \textit{Gaussian approximation} of the bispectrum covariance.}
  {We study the range of validity of this Gaussian approximation for two-dimensional non-Gaussian random fields.}
  {For this purpose, we simulate Gaussian and non-Gaussian random fields, the latter represented by log-normal fields and obtained directly from the former by a simple transformation. From the simulated fields, we calculate the power spectra, the bispectra, and the covariance from the sample variance of the bispectra, for different degrees of non-Gaussianity $\alpha$, which is equivalent to the skewness on a given angular scale $\theta_{\rm g}$. In doing so, we minimize sample variance by selecting the same `phases' of the Gaussian and non-Gaussian fields.}
  {We find that the Gaussian approximation of the bispectrum covariance provides a good approximation for degrees of non-Gaussianity $\alpha\le 0.5$ and a reasonably accurate approximation for $\alpha\le 1$, both on scales $\gtrsim8\theta_{\rm g}$. For larger values of $\alpha$, the Gaussian approximation clearly breaks down. Using results from cosmic shear simulations, we estimate that the cosmic shear convergence fields are described by $\alpha\lesssim0.7$ at $\theta_{\rm g}\sim4^{\prime\prime}$.}
  {We therefore conclude that the Gaussian approximation for the bispectrum covariance is likely to be applicable in ongoing and future cosmic shear studies.}


  \maketitle


\section{Introduction}
\label{sec:Introduction}
Huge cosmological surveys are currently ongoing or planned for the near future. The large statistical power of these surveys allows us to study the statistical properties of the cosmological fields well beyond the lowest, second-order statistical level. In order to employ these higher-order statistics in obtaining information about the underlying cosmology, one needs to be able to accurately predict these statistics as a function of the model parameters. Moreover, one also needs to predict the expected errors on these higher-order measures, i.e., the covariance of higher-order statistics.

Whereas there exist rather accurate fitting formulae for the power spectrum of the cosmic density field \citep[e.g.,][]{1996MNRAS.280L..19P, 2003MNRAS.341.1311S}, this is no longer the case for higher-order statistical properties. For example, the semi-empirical fitting formula for the bispectrum of the cosmic density field \citep{2001MNRAS.325.1312S} is estimated to be accurate only at the level of several tens of percent.

In order to obtain the covariance of the power spectrum and the bispectrum, even higher-order statistical properties of the field need to be known. The power spectrum covariance depends on fourth-order statistical properties, i.e., the trispectrum, which is even harder to predict. In principle, one needs to understand the sixth-order properties of the field in order to determine the covariance of the bispectrum. 

However, depending on the statistical field under consideration, certain simplifications may apply. In particular, if the density field is sufficiently close to a Gaussian field, the higher-order terms in the covariance of the power spectrum and the bispectrum may become subdominant, in which case they can be obtained solely from the power spectrum itself. 

In this paper we study the covariance of the bispectrum for two-dimensional random fields. Whereas our prime interest is focused on the convergence field of cosmic shear (as a prime example of a two-dimensional cosmological field) \citep[for a review of gravitational lensing, see][]{2006glsw.conf....1S}, our results are not restricted to this particular application.

Third-order cosmic shear statistics is known to provide very valuable information, complementary to that from the cosmic shear power spectrum \citep[e.g.,][]{2004MNRAS.348..897T, 2005A&A...442...69K, 2006MNRAS.366..101H}, which is confirmed by the first measurements of third-order cosmic shear in previous surveys \citep{2002A&A...389L..28B, 2003ApJ...592..664P, 2004MNRAS.352..338J, 2011MNRAS.410..143S}. Ongoing and future surveys, such as the Canada-France-Hawaii-Telescope Legacy Survey (CFHTLS), the KIlo-Degree Survey (KIDS), the Dark Energy Survey (DES) and EUCLID, will measure the third-order cosmic shear signal with very high accuracy. However, estimating the statistical power of these surveys, and obtaining realistic confidence estimates for model parameters, remains a challenge that hinges on our ability to estimate the bispectrum covariance.

\citet{2000PhRvD..62d3007H} estimated this covariance by analogy to the bispectrum covariance of the cosmic microwave background. As discussed in \citet[][hereafter JSS]{2009A&A...508.1193J}, this approach has a number of drawbacks that result from the use of spherical harmonics in the analysis. Instead, JSS derived a general expression for the bispectrum covariance on a finite region in the flat-sky limit, which can be separated into four terms. The first term contains the contribution of the power spectrum to the bispectrum covariance, and is the only term relevant for Gaussian random fields. The other three terms contain the contributions of the bispectrum, the trispectrum and the pentaspectrum. Dropping these higher-order terms in the expression is usually termed the Gaussian approximation of the bispectrum covariance. If the other terms were small for realistic cosmic fields, this would tremendously simplify the estimation of confidence regions from third-order statistics. At first glance this seems to be invalid as the present density field appears to be highly non-Gaussian, especially on small scales. Nevertheless, this Gaussian approximation has been employed in nearly all studies of parameter forecasts from third-order cosmic shear \citep[e.g.,][]{2004MNRAS.348..897T, 2006MNRAS.366..101H, 2010A&A...523A..60S}.

In this paper, we investigate the validity of the Gaussian approximation, using simulated random fields with different degrees of non-Gaussianity. Specifically, we employ log-normal random fields which we characterise by the non-Gaussianity parameter $\alpha$. As argued in the seminal paper by \citet{1991MNRAS.248....1C}, matter density fields are well described by log-normal statistics, making these statistics relevant in a cosmological context. Indeed, as \citet{2001ApJ...561...22K} have shown, the distribution of the density fluctuations in the universe can rather well be described by a log-normal distribution. Moreover, \citet{2002ApJ...571..638T} and \citet{2011arXiv1105.3980H} have shown that the same holds true for the cosmic shear convergence field, which serves as justification for using these particular non-Gaussian fields for our study. Furthermore, \citet{2011arXiv1105.3980H} employed the log-normal ansatz to calculate the covariance of the power spectrum of cosmic shear fields; they showed that this model provides fairly accurate results, when compared to ray-tracing simulations. We investigate the question to which degree of non-Gaussianity $\alpha$ it is justified to use the Gaussian approximation for the bispectrum covariance.

This paper is organised as follows: Sect.~2 reviews the most important properties of the analytical formula for the bispectrum covariance derived by JSS and the corresponding Gaussian approximation, as well as an ansatz to determine the scope of validity of the latter applied to non-Gaussian fields. In Sect.~3 the primary quantities in the theoretical framework of log-normal random fields are introduced, and the method we used to numerically generate the random fields. The results will be presented, analysed and discussed in Sect.~4. Finally, Sect.~5 provides conclusions.


\section{The bispectrum covariance: Analytical formula and Gaussian approximation}
\label{sec:The bispectrum covariance - analytical formula and approximation}
In general, the bispectrum of a two-dimensional homogeneous and isotropic random field in Fourier space $f(\vec{\ell})$ with zero mean is defined by
\begin{equation}
 \left\langle f(\vec{\ell}_1)\;f(\vec{\ell}_2)\;f(\vec{\ell}_3)\right\rangle = (2\pi)^2\; \delta^{(2)}_{\rm D}(\vec{\ell}_1+\vec{\ell}_2+\vec{\ell}_3)\; B(\ell_1,\ell_2,\ell_3)\,,
 \label{eq:bi}
\end{equation}
and consequently depends on the length $\ell_i=|\vec{\ell}_i|$ of the three angular frequency vectors $\vec{\ell}_1,\vec{\ell}_2,\vec{\ell}_3$. It is characterised by the fact that the two-dimensional Dirac delta-distribution $\delta^{(2)}_{\rm D}(\vec{\ell})$ restricts the three angular frequency vectors to form a triangle. Accordingly, JSS introduced the following bispectrum estimator for a random field of size $\Omega$,
\begin{eqnarray}
 \hat{B}(\bar{\ell}_{1},\bar{\ell}_{2},\bar{\ell}_{3}) &:=& \frac{(2\pi)^{2}}{\Omega} \Lambda^{-1}\left(\bar{\ell}_{1},\bar{\ell}_{2},\bar{\ell}_{3}\right) \nonumber\\ 
 & & \times \left[\prod_{i=1}^{3} \int_{A_{R}(\bar{\ell}_{i})} \frac{{\rm d}^{2} \ell_{i}}{A_{R}(\bar{\ell}_{i})} f(\vec{\ell}_{i})\right] \delta^{(2)}_{\rm D}(\vec{\ell}_{1}+\vec{\ell}_{2}+\vec{\ell}_{3})\,, 
 \label{eq:biesti}
\end{eqnarray}
so that the bispectrum can be calculated by averaging configurations over annuli. The mean length of a triangle side, corresponding to the mean radius of an annulus, is denoted by $\bar{\ell_i}$ and the corresponding area of an annulus is given by $A_{R}(\bar{\ell_i})=2\pi \bar{\ell_i} \Delta {\ell_i}$, with $\Delta {\ell_i}$ being the respective bin size. The parameter 
\begin{equation}
 \Lambda^{-1} \left(\bar{\ell_1},\bar{\ell_2},\bar{\ell_3}\right) = \frac{1}{4} \sqrt{2\bar\ell^{2}_{1}\bar\ell^{2}_{2} + 2\bar\ell^{2}_{1}\bar\ell^{2}_{3} + 2\bar\ell^{2}_{2}\bar\ell^{2}_{3} - \bar\ell^{4}_{1} - \bar\ell^{4}_{2} - \bar\ell^{4}_{3}}\,
\end{equation}
reflects the area of the triangle which is spanned by the three wave vectors. JSS showed that the estimator (\ref{eq:biesti}) is unbiased in the limit of small annuli width $\Delta \ell_i$.

JSS derived an analytical expression for the covariance of the bispectrum estimator defined in Eq.~(\ref{eq:biesti}),
\begin{eqnarray}
 {\rm Cov} \left(\hat B(\bar{\ell}_{1},\bar{\ell}_2,\bar{\ell}_3),\,\hat B(\bar{\ell}_4,\bar{\ell}_5,\bar{\ell}_6)\right) 
 &=& \gamma\;P(\bar{\ell}_{1})P(\bar{\ell}_{2})P(\bar{\ell}_{3})\nonumber\\
 &+& T_{{\tens 3} \times {\tens 3}} +T_{{\tens 4} \times {\tens 2}} + T_{\tens 6}\,.
\label{eq:bc}
\end{eqnarray}
This expression consists of four different terms. The first term is proportional to the third power of the power spectrum, with an amplitude given by the geometrical factor 
\begin{equation}
 \gamma := \frac{(2\pi)^{3}}{\Omega\; \bar{\ell}_{1}\bar{\ell}_{2}\bar{\ell}_{3} \Delta \ell_{1}\Delta \ell_{2}\Delta \ell_{3}} \;\Lambda^{-1}\left(\bar{\ell}_{1},\bar{\ell}_{2},\bar{\ell}_{3}\right) \;D_{\bar{\ell}_{1},\bar{\ell}_{2},\bar{\ell}_{3},\bar{\ell}_{4},\bar{\ell}_{5},\bar{\ell}_{6}}\,,
\label{eq:gamma}
\end{equation}
with the short hand notation
\begin{eqnarray}
 D_{\bar{\ell}_{1},\bar{\ell}_{2},\bar{\ell}_{3},\bar{\ell}_{4},\bar{\ell}_{5},\bar{\ell}_{6}} 
 &=& \delta_{\bar\ell_{1}\bar\ell_{4}}\delta_{\bar\ell_{2}\bar\ell_{5}}\delta_{\bar\ell_{3}\bar\ell_{6}} + \delta_{\bar\ell_{1}\bar\ell_{5}}\delta_{\bar\ell_{2}\bar\ell_{4}} \delta_{\bar\ell_{3}\bar\ell_{6}} \nonumber\\
 &+& \delta_{\bar\ell_{1}\bar\ell_{4}}\delta_{\bar\ell_{2}\bar\ell_{6}}\delta_{\ell_{3}\ell_{5}} + \delta_{\bar\ell_{1}\bar\ell_{5}}\delta_{\bar\ell_{2}\bar\ell_{6}}\delta_{\bar\ell_{3}\bar\ell_{4}} \nonumber\\
 &+& \delta_{\bar\ell_{1}\bar\ell_{6}}\delta_{\bar\ell_{2}\bar\ell_{4}}\delta_{\bar\ell_{3}\bar\ell_{5}} + \delta_{\bar\ell_{1}\bar\ell_{6}}\delta_{\bar\ell_{2}\bar\ell_{5}}\delta_{\bar\ell_{3}\bar\ell_{4}} \:,
\end{eqnarray}
and $\delta_{xy}=1$ if $x=y$, and zero otherwise. In the case of a Gaussian field, this would be the sole contribution to the covariance; therefore, the approximation of the bispectrum covariance by this first term only is called \textit{Gaussian approximation} in the following. The other three terms account for the contribution of higher-order spectra to the bispectrum covariance: The second term is an integral over the product of bispectra, the third term an integral over the product of the trispectrum and the power spectrum, and finally the fourth term an integral over the pentaspectrum.

Due to the complexity of Eq.~(\ref{eq:bc}) and the fact that higher-order spectra are usually much more difficult to obtain, either analytically or from simulations, usually only the Gaussian approximation is used for statistical analyses of cosmological fields, despite the fact that these fields are not purely Gaussian (see Sect.~\ref{sec:Introduction}). Thus the question arises to which degree of non-Gaussianity of the underlying field it is justified to use the Gaussian approximation for computing the bispectrum covariance. We have addressed the problem by the following approach: Applied to a purely Gaussian random field, the Gaussian approximation should hold per definition,
\begin{equation} 
 {\rm Cov}_{\rm G} \left(\hat B(\bar{\ell}_{1},\bar{\ell}_2,\bar{\ell}_3),\, \hat B(\bar{\ell}_4,\bar{\ell}_5,\bar{\ell}_6)\right) 
 = \gamma\;P_{\rm G}(\bar{\ell}_{1})P_{\rm G}(\bar{\ell}_{2})P_{\rm G}(\bar{\ell}_{3})\,,
 \label{eq:app}
\end{equation}
as will nevertheless be investigated numerically in Sect.~\ref{subsubsec:Accuracy for the Gaussian case}. However, for a non-Gaussian random field, the Gaussian approximation might hold to a certain degree of non-Gaussianity $\alpha$ (which will be specified in Sect.~\ref{subsec:Log-normal random fields}) as well,
\begin{equation}
 {\rm Cov}_{\alpha}\left(\hat B(\bar{\ell}_{1},\bar{\ell}_{2},\bar{\ell}_{3}),\,\hat B(\bar{\ell}_{4},\bar{\ell}_{5},\bar{\ell}_{6})\right) 
 \approx \gamma\;P_{\alpha}(\bar{\ell}_{1})P_{\alpha}(\bar{\ell}_{2})P_{\alpha}(\bar{\ell}_{3})\,.
\label{eq:applog}
\end{equation}
In order to assess the range of validity of the Gaussian approximation for non-Gaussian random fields, we can compare a Gaussian random field with a non-Gaussian one, characterised by the degree of non-Gaussianity $\alpha$, both having the same field geometry to assure the same geometrical factor $\gamma$ for both fields. The quality of the Gaussian approximation can then be quantified, on the one hand, by a parameter $\varsigma$, which is defined by
\begin{eqnarray}
 \varsigma(\alpha) &=& \frac{{\rm Cov}_{\alpha}\left(\hat B(\bar{\ell}_{1},\bar{\ell}_{2},\bar{\ell}_{3}),\,\hat B(\bar{\ell}_{1},\bar{\ell}_{2},\bar{\ell}_{3})\right)}{P_{\alpha}(\bar{\ell}_{1})P_{\alpha}(\bar{\ell}_{2})P_{\alpha}(\bar{\ell}_{3})}\nonumber\\
 &\times& \left[\frac{{\rm Cov}_{\rm G}\left(\hat B(\bar{\ell}_{1},\bar{\ell}_{2},\bar{\ell}_{3}),\hat B(\bar{\ell}_{1},\bar{\ell}_{2},\bar{\ell}_{3})\right)}{P_{\rm G}(\bar{\ell}_{1})P_{\rm G}(\bar{\ell}_{2})P_{\rm G}(\bar{\ell}_{3})}\right]^{-1}.
\label{eq:ansatz}
\end{eqnarray}
The first term on the right side corresponds to the geometrical factor $\gamma$ defined in Eq.~(\ref{eq:applog}), the second term to $\gamma^{-1}$ defined in Eq.~(\ref{eq:app}). As long as these to factors give the same result, $\varsigma(\alpha) \approx 1$ and the Gaussian approximation provides a reasonable description of the diagonal part of the bispectrum covariance. On the other hand, one needs to check whether the true covariance matrix is (close to) diagonal, as predicted in the Gaussian approximation.

\section{Methods} 
\label{sec:Methods}

\subsection{Log-normal random fields} 
\label{subsec:Log-normal random fields}
\begin{figure}[t]
 \resizebox{\hsize}{!}{\includegraphics{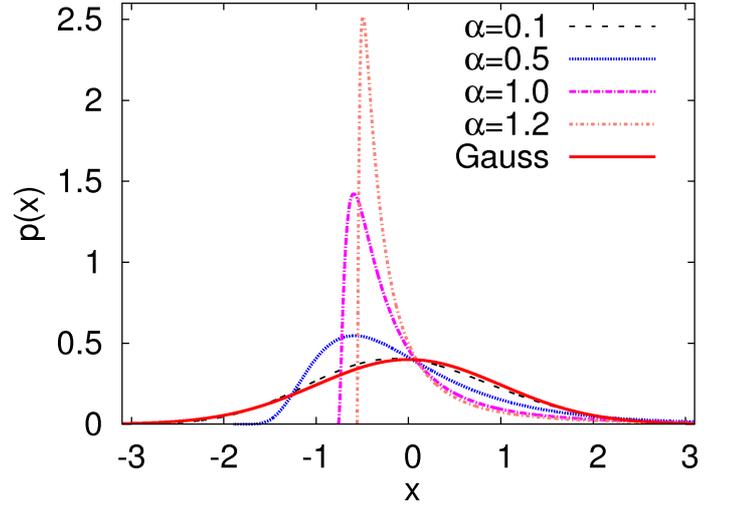}}
 \caption{Gaussian and log-normal distribution with zero mean and unit variance, the latter for different $\alpha$. The larger $\alpha$ is, the stronger the log-normal distribution deviates from the Gaussian.}
 \label{fig:log-normaldis}
\end{figure}

To estimate the parameter $\varsigma(\alpha)$, we consider a particular case of a non-Gaussian field, namely a log-normal field. This is motivated by its relative simplicity, as well as the fact that the density fluctuations in cosmological models can reasonably well be approximated by a log-normal distribution (see Sect.~\ref{sec:Introduction}). The transformation of a two-dimensional homogeneous and isotropic Gaussian random field $g(\vec{x})$ with zero mean and unit variance to a homogeneous and isotropic two-dimensional log-normal random field $L(\vec{x})$ is given by
\begin{equation}
 L(\vec{x}) = \frac{A}{c} \, \left({\rm e}^{\alpha\, g(\vec{x})} -c\right)\;.
 \label{eq:trafo}
\end{equation}
The parameter $\alpha$ controls the deviation of the log-normal field from the initial Gaussian field and thus the degree of non-Gaussianity of the former. The parameter $c$ and $A$ are chosen as such that the log-normal field has zero mean and unit variance,
\begin{eqnarray}
 \mu_{\alpha} 
 &=& \left\langle L(\vec{x}) \right\rangle = \int_{-\infty}^{\infty} {\rm d}g(\vec{x})~L(\vec{x})~p(g(\vec{x})) = \frac{A}{c} \left({\rm e}^{\alpha^2/2} - c\right) = 0 \nonumber\\
 &\Rightarrow& c= {\rm e}^{\alpha^2/2}\;,
\end{eqnarray}
\begin{eqnarray}
 \sigma^{2}_{\alpha} &=& \left\langle L^{2}(\vec{x})\right\rangle = \int_{-\infty}^{\infty} {\rm d}g(\vec{x}) ~ L^{2}(\vec{x}) \, p(g(\vec{x})) = A^{2} (c^{2}-1) =1 \nonumber\\
 &\Rightarrow& A =\frac{1}{\sqrt{c^{2}-1}}=\frac{1}{\sqrt{{\rm e}^{\alpha^2}-1}} \,.
 \label{eq:logvariance}
\end{eqnarray}
Obviously, $c$ and $A$ depend only on the deviation parameter $\alpha$ so that Eq.~(\ref{eq:trafo}) can be reduced in its dependence on $\alpha$,
\begin{equation}
 L(\vec{x}) = \frac{1}{{\rm e}^{\alpha^2/2}\sqrt{{\rm e}^{\alpha^{2}}-1}} \, \left({\rm e}^{\alpha\, g(\vec{x})} -{\rm e}^{\alpha^2/2}\right)\;.
 \label{eq:Log-normal}
\end{equation}
The probability density of the Gaussian random field $g$ is a Gaussian with zero mean and unit variance, and the one of the log-normal random field $L$ can be obtained from the Gaussian by the relation $p\left(g\right) {\rm d} g=p_{\alpha}\left(L\right){\rm d} L$, yielding
\begin{equation}
 p_{\alpha}\left(L\right)
 = \frac{1}{\sqrt{2\pi} \alpha (L +A)}\,{\rm exp}\left( -\frac{\left[{\rm ln}\left(\frac{L}{A} +1\right) + \frac{\alpha^{2}}{2}\right]^{2}}{2\alpha^{2}} \right) \;.
 \label{eq:pdln}
\end{equation}
The impact of the parameter $\alpha$ is visualised in Fig.~\ref{fig:log-normaldis}, which compares the Gaussian probability distribution with the log-normal probability distribution for different values of $\alpha$. $p_\alpha(L)$ peaks at negative $L$ (for $\alpha>0$) and shows an extended tail for positive $L$. With increasing $\alpha$, these features become more pronounced. Furthermore, the probability distribution vanishes for $L<L_{\rm min}=-A$, as can be easily seen from Eq.~(\ref{eq:trafo}).

\subsection{Numerical generation of the random fields}
\label{subsec:Numerical generation of the random fields}
In order to evaluate the parameter $\varsigma(\alpha)$, we need to calculate the bispectrum covariance of the Gaussian and the log-normal field, as well as the power spectra. This has been done by numerical simulations: Using repeated realisations of the random fields, we can determine their power spectra and bispectra, as well as the covariance of the bispectra from the sample variance among these different realisations. More specifically, we generated realisations of a Gaussian random field with a predefined power spectrum (see below)
which uniquely determines its properties. For each realisation of the Gaussian field, we obtained one realisation of a log-normal field applying the transformation (\ref{eq:trafo}) to every pixel, for each chosen value of $\alpha$. The Gaussian field power spectrum is normalised such that the variance on the pixel scale is unity. In this way, the geometries of the Gaussian and log-normal fields are the same, and the phases of their realisations are the same, which guarantees to minimize sample variance in the comparison carried out in Eq.~(\ref{eq:ansatz}).

A two-dimensional Gaussian random field can be generated easiest in Fourier space using the standard procedure, i.e. assuming periodicity and assigning Fourier coefficients with mean squared amplitude
\begin{equation}
 \langle |g_{\vec{\ell}}|^{2} \rangle = \left(\frac{\Delta \ell}{2\pi}\right)^{2} P(|\vec{\ell}|) = \sigma^{2}_{\rm G}(|\vec{\ell}|)\;. 
 \label{eq:Gaussianfield}
\end{equation}
The variance $\sigma^{2}_{\rm G}(|\vec{\ell}|)$ of the Gaussian random field $\tilde g(\vec{\ell})$ is uniquely determined by the choice of an input power spectrum $P(|\vec{\ell}|)$ and the choice of a smallest wave number $\ell_{\rm min}$, or equivalently, the field size $\Omega$. It is needed as input for a Gaussian random number generator \citep[see][]{Gnu}, which draws separately the real and imaginary parts of $g_{\vec{\ell}}$ for each wave vector $\vec{\ell}$ from a Gaussian distribution with standard deviation $\sigma_{\rm G}(|\vec{\ell}|)\propto \sqrt{P(|\vec{\ell}|)}$. For our purpose, we chose the input power spectrum to be of the form 
\begin{equation}
 P(\ell)\propto\ell(\ell+\ell_{0})^{-4}\;,
 \label{eq:powerspectrum}
\end{equation}
with a normalisation that leads to a unit variance of $g$, and we put $\ell_{0}=1000$. The size of the field is fixed to $\Omega=10^\circ \times 10^\circ$, leading to a smallest wave number of $\ell_{\rm min} \approx 36$. The field is discretised onto $N^2=1024^2$ grid points.

As the log-normal field is obtained from the corresponding Gaussian field by a transformation in real space (see Eq.~\ref{eq:trafo}), the Gaussian field is transformed from Fourier space to real space using the Fast Fourier Transform algorithm FFTW \citep[see][]{FFTW05}. After the log-normal field is calculated according to Eq.~(\ref{eq:trafo}), it is then transformed back to Fourier space.

To achieve a good signal-to-noise for the comparison of the bispectrum covariance matrix of the Gaussian with those of the log-normal random fields, it is crucial to generate a large number $N_{\rm r}$ of realisations of the corresponding field; we chose $N_{\rm r}=1000$. For each field the power spectrum and the bispectrum are computed for each of these $N_{\rm r}$ realisations and then the average over these $N_{\rm r}$ power spectra and bispectra is calculated, as well as the variance of the bispectra, which yields an estimate of the bispectrum covariance.


\section{Results}
\label{sec:Results} 

\subsection{Power spectrum}
\label{subsec:Power spectrum} 
\begin{figure}[b]
 \resizebox{\hsize}{!}{\includegraphics{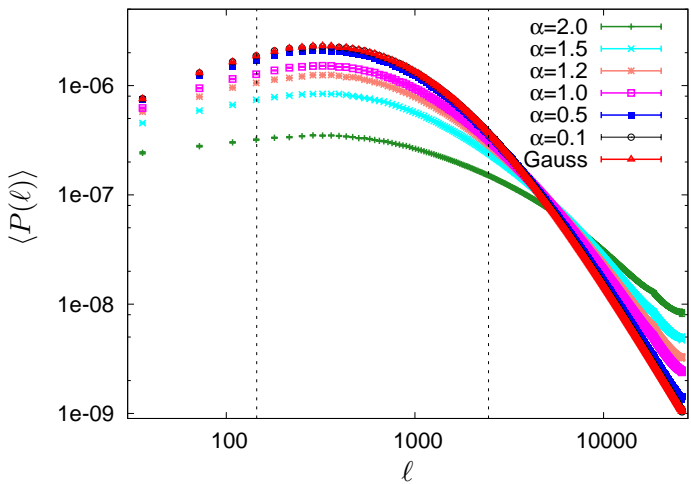}}
 \caption{The mean power spectrum, averaged over $1000$ realisations, of the Gaussian random field and the mean log-normal power spectrum for different degrees of non-Gaussianity $\alpha$. The error bars are the standard deviation of the $1000$ power spectra divided by $\sqrt{1000}$.}
 \label{fig:po}
\end{figure}
To determine the range of validity of the Gaussian approximation according to Eq.~(\ref{eq:ansatz}), the measurement of the power spectrum of the Gaussian and of the log-normal fields is necessary. The power spectrum is computed for every realisation of each field as the average of the Fourier coefficients $|g_{\vec{\ell}}|^{2}$ and $|L_{\vec{\ell}}|^{2}$, respectively, over the polar angle of $\vec\ell$. 

Fig.~\ref{fig:po} shows the mean power spectrum obtained by averaging over the $1000$ power spectra for each field in the range $\ell\approx 36 - 26000$. To determine the bispectrum covariance, the power spectrum and bispectrum are measured in bins equidistant with $\Delta\ell =110$ in the range highlighted in Fig.~\ref{fig:po} ($\ell=145 - 2455$). In the following, we restrict all measurements to the range $\ell=145 - 2455$, since for substantially larger $\ell$, the computing time for the bispectrum becomes quite long.

The power spectra of the Gaussian and the log-normal random field do not differ noticeably for $\alpha=0.1$; in fact, these two cannot be distinguished in Fig.~\ref{fig:po}. However, for $\alpha\geq0.5$, the power spectrum of the log-normal field starts to differ from that of the Gaussian. Given that the log-normal transformation has positive curvature, it increases the amplitude and decreases the width of density peaks. This leads to an increase of power on small scale, as is clearly seen in Fig.~\ref{fig:po}. Since the variances of all fields are the same on the pixel scale, this increase of power for large $\ell$ must be compensated by a corresponding decrease of the power for large scales.

\subsection{Bispectrum}
\label{Bispectrum}
According to Eq.~(\ref{eq:bi}), one way to compute the bispectrum of a random field is to calculate the 3-point correlator in Fourier space and to normalise it afterwards by the corresponding prefactor. We employed such a code, which generates at the beginning of the computational process a list of all Fourier coefficients contained in the different bins defined in Sect.~\ref{subsec:Power spectrum}. For every bin, the code loops over all combinations of the wave vectors $\vec{\ell}_{1}$, $\vec{\ell}_{2}$ and $\vec{\ell}_{3}$ and checks whether the triangle condition is satisfied, i.e., $\vec{\ell}_{1}+\vec{\ell}_{2}+\vec{\ell}_{3}=0$. The wave vectors are further constrained to form triangles close to the targeted shape and size, with an allowed deviation of $\Delta \ell_{i}=55$.

\begin{figure}[b]
 \resizebox{\hsize}{!}{\includegraphics{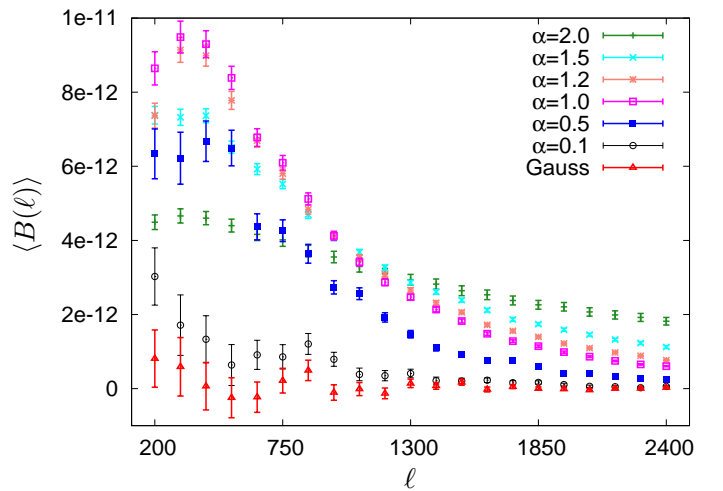}}
 \caption{The mean bispectrum per bin for equilateral triangles, averaged over $1000$ realisations, of the Gaussian random field and several log-normal fields with different values of the non-Gaussianity parameter $\alpha$. The error bars are the standard deviation of the $1000$ power spectra divided by $\sqrt{1000}$.}
 \label{fig:bi}
\end{figure} 

We first restrict our consideration to equilateral triangles, i.e., that all $\bar \ell_i$ are equal. The mean bispectrum of the $1000$ realisations of the Gaussian and log-normal random fields for different values of $\alpha$ is shown in Fig.~\ref{fig:bi}. As expected, the bispectrum of the Gaussian field is compatible with zero within the statistical accuracy of our simulations. The size of the error bars increases towards smaller $\ell$ due to the smaller number of modes (or available triangles) per $\ell$-bin. For the log-normal fields, the bispectrum deviates significantly from zero even for $\alpha=0.1$. For small values of $\alpha$ the amplitude of the bispectrum increases with $\alpha$ at all values of $\ell$ considered. As seen in Fig.~\ref{fig:po}, for $\alpha\lesssim 0.5$, the power spectra of the log-normal fields are very similar, and therefore the increase of the bispectrum with $\alpha$ shows an increasing degree of non-Gaussianity. For larger values of $\alpha$, the bispectrum no longer behaves monotonically with $\alpha$ at all $\ell$, which is related to the dependence of the power spectrum on $\alpha$. To leading order, the third-order statistics of a log-normal field is the product of an $\alpha$-dependent factor describing the degree of non-Gaussianity, and the square of second-order statistics. Hence, the decrease of the bispectrum with $\alpha$ for larger values of $\alpha$ and smaller $\ell$ reflects the decrease of the power spectrum with $\alpha$ for these $\ell$, which results from our normalisation condition. For large $\ell$, the bispectrum monotonically increases with $\alpha$.

\subsubsection{The reduced bispectrum - Estimating the degree of non-Gaussianity}
\label{subsubsec:The reduced bispectrum - Estimating the degree of non-Gaussianity}
\begin{figure}[t]
 \resizebox{\hsize}{!}{\includegraphics{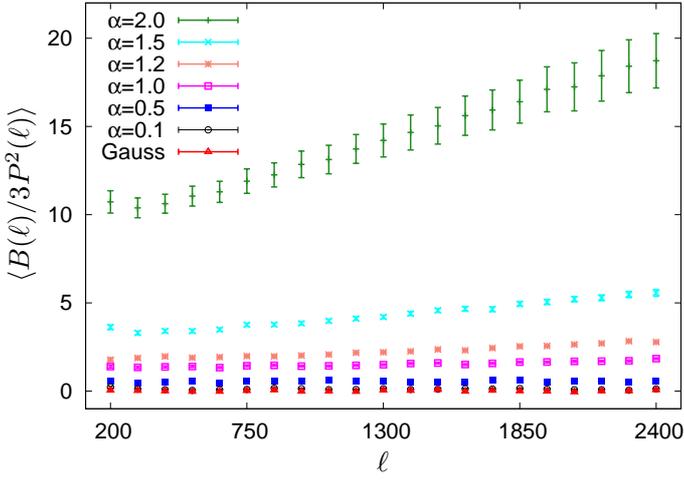}}
 \caption{The reduced bispectrum $B(\ell) / 3P^{2}(\ell)$ for a Gaussian and several log-normal random fields with different $\alpha$. The error bars are calculated by Gaussian error propagation.}
 \label{fig:nongauss}
\end{figure}

In order to remove the leading, quadratic dependence of the bispectrum on the power spectrum, we consider the reduced bispectrum, defined as 
\[B(\ell_1,\ell_2,\ell_3) [P(\ell_1)P(\ell_2)+P(\ell_1)P(\ell_3)+P(\ell_2)P(\ell_3)]^{-1}\;,\]
which for equilateral triangles reduces to $B(\ell) / 3P^{2}(\ell)$. This reduced bispectrum is shown for different $\alpha$ in Fig.~\ref{fig:nongauss}. As expected, its amplitude grows with increasing $\alpha$; for $\alpha=1.5$ it is approximately five times larger than for $\alpha=0.1$. Additionally, the reduced bispectrum increases with $\ell$, indicating a larger degree of non-Gaussianity on smaller scales.

\subsection{Bispectrum covariance}
\label{sub:Bispectrum covariance}

\subsubsection{Accuracy for the Gaussian case}
\label{subsubsec:Accuracy for the Gaussian case}
\begin{figure}[t]
 \resizebox{\hsize}{!}{\includegraphics{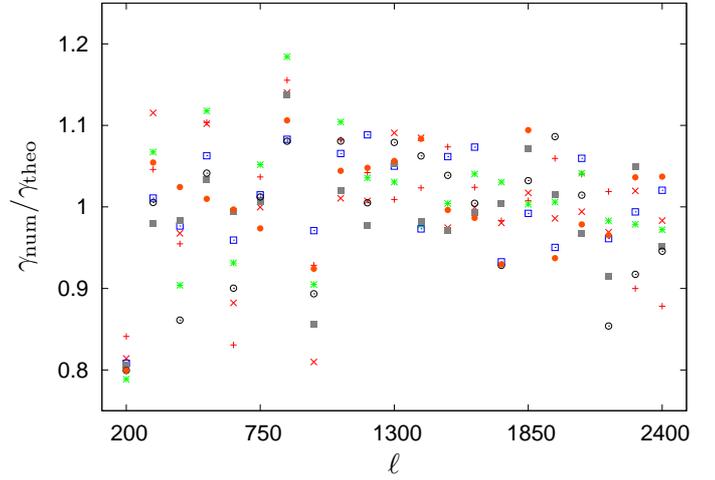}} 
 \caption{The relation between the theoretically calculated geometrical factor and the numerically computed geometrical factor $\gamma_{\rm num} / \gamma_{\rm theo}$ for seven sets of 1000 realisations of a Gaussian random field. For clarity the error bars are not plotted.}
 \label{fig:gammarel}
\end{figure}

Before testing the range of validity of the Gaussian approximation for the bispectrum covariance of non-Gaussian fields, it is reasonable to first probe the accuracy of the Gaussian approximation for a purely Gaussian random field. For this, we calculate the ratio of the numerically determined bispectrum covariance and third-order products of the power spectrum, to obtain the geometrical factor $\gamma_{\rm num}$ from the simulations, which is then compared to the expected value $\gamma_{\rm theo}$ given by Eq.~(\ref{eq:gamma}). The value of $\gamma_{\rm num}$ has been determined for seven sets of $1000$ realisations individually; the ratio $\gamma_{\rm num}/\gamma_{\rm theo}$ for these seven sets is shown in Fig.~\ref{fig:gammarel}. The ratio fluctuates around one, with higher fluctuations for smaller wave numbers $\ell$, which is due to the fact that the results on large scales are obtained by averaging over fewer triangles than on small scales. Only at the smallest value of $\ell$ considered do we see a significant deviation of this ratio from unity. This deviation is not surprising, given the various approximations that were made for deriving Eq.~(\ref{eq:bc}), the most important of which is that the spatial scale is much smaller than the size of the field, so that edge
effects can be neglected. For $\ell=200$, corresponding to about $1\thinspace{\rm deg}$, this assumption is no longer fully justified. However, for larger values of $\ell$ we conclude that the analytical expression for the geometrical factor is accurate.

\subsubsection{Diagonal elements}
\label{subsubsec:Diagonal elements}
\begin{figure}[t]
 \resizebox{\hsize}{!}{\includegraphics{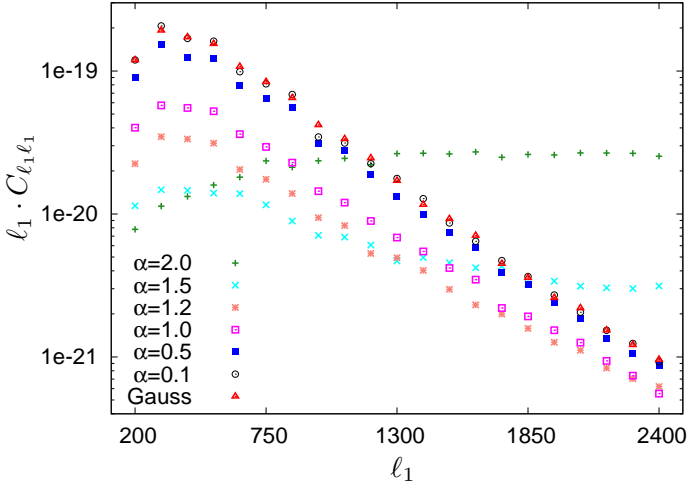}}
 \caption{The diagonal elements ($\ell_{1}=\ell_{2}$) of the Gaussian covariance matrix compared to the diagonal elements of the log-normal covariance matrix for different degrees of non-Gaussianity $\alpha$. For clarity, the diagonal elements are multiplied by the factor $\ell_{1}$, and the error bars are not plotted.}
 \label{fig:dia}
\end{figure}
The Gaussian approximation for the bispectrum covariance says that for equilateral triangles,
\begin{equation}
 C_{\bar\ell_1\bar\ell_2}
 \equiv {\rm Cov} \left( \hat B(\bar\ell_1,\bar\ell_1,\bar\ell_1),\, \hat B(\bar\ell_2,\bar\ell_2,\bar\ell_2) \right) =\gamma P^3(\bar\ell_1) \delta_{\bar \ell_1 \bar\ell_2}\;,
\end{equation}
i.e., that the off-diagonal elements of $C_{\ell_1\ell_2}$ are zero, and that the diagonal elements follow the Gaussian behaviour. Here, we first study the latter, and show the diagonal elements in Fig.~\ref{fig:dia}. For $\alpha\lesssim 0.5$, the diagonal elements for the log-normal case show no significant difference from the Gaussian case, whereas for $\alpha\gtrsim 0.5$ the amplitude of the diagonal elements of the log-normal covariance matrix decreases in comparison to the Gaussian covariance matrix, reflecting the decrease of the power spectrum with $\alpha$ across the $\ell$ range considered. For large values of $\alpha$, $C_{\ell \ell}$ increases again, due to the larger degree of non-Gaussianity, where the higher-order terms of Eq.~(\ref{eq:bc}), unaccounted for in the Gaussian approximation, start dominating the behaviour of the covariance matrix.

\begin{figure}[b]
 \resizebox{\hsize}{!}{\includegraphics{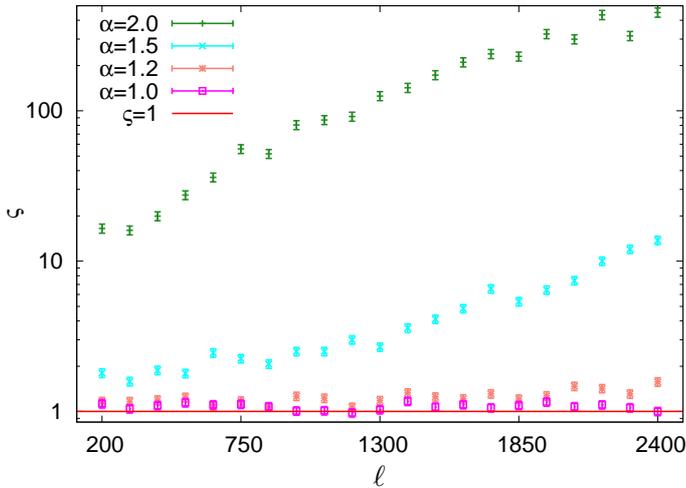}} 
 \caption{Aberration parameter $\varsigma$ for large degrees of non-Gaussianity $\alpha$. The error bars are computed by Gaussian error propagation using $\Delta {\rm Cov} = {C}_{ij}\;\sqrt{2/(1000-1)}$.}
 \label{fig:relstand}
\end{figure}

\begin{figure}[t]
 \resizebox{\hsize}{!}{\includegraphics{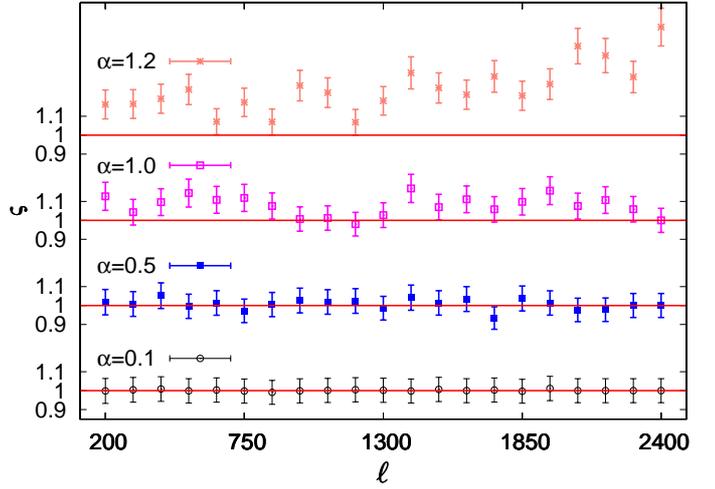}} 
 \caption{Aberration parameter $\varsigma$ for small degrees of non-Gaussianity $\alpha$.}
 \label{fig:relstandzoom}
\end{figure}

To remove the dependence of the covariance from the power spectrum, we consider the aberration parameter $\varsigma(\alpha)$ defined in Eq.~(\ref{eq:ansatz}). This is shown in Fig.~\ref{fig:relstand} for several relatively large values of $\alpha$. For this figure, we study the aberration parameter down to roughly eight times the pixel scale. This figure indicates that $\varsigma$ is of order unity for $\alpha\lesssim 1.2$, but strongly increases for larger values of $\alpha$, indicating a clear breakdown of the Gaussian approximation. To study in more detail the transition to this breakdown, we show in Fig.~\ref{fig:relstandzoom} the behaviour of $\varsigma$ with $\ell$ for several smaller values of $\alpha$. This figure indicates that for $\alpha=0.5$, $\varsigma\approx 1$ within the statistical accuracy, and that $\varsigma\approx 1.1$ for $\alpha=1.0$. For $\alpha=1.2$, the deviations are already several tens of percent, and clearly increase towards larger $\ell$. Hence we conclude that the diagonal elements of the covariance are well described by the Gaussian approximation for $\alpha\lesssim 0.5$, on the scales probed, and that the deviation from the Gaussian approximation are confined to less than 10\% for $\alpha\lesssim 1$.

\subsubsection{Off-diagonal elements}
\label{subsubsec:Off-diagonal elements}
\begin{figure}[t]
 \resizebox{\hsize}{!}{\includegraphics{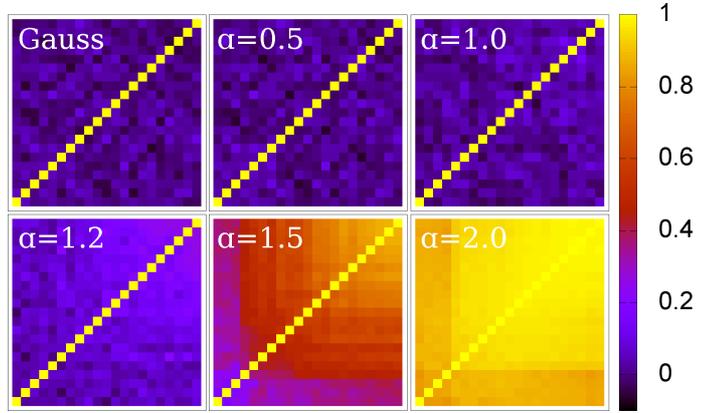}} 
 \caption{Gaussian and log-normal correlation matrices for different $\alpha$. The wave numbers $\ell_{1}$ and $\ell_{2}$ rise for each panel from the left lower corner to the right  upper corner.}
 \label{fig:offdia}
\end{figure}

\begin{figure}[t]
 \sidecaption
 \includegraphics[width=4.4cm]{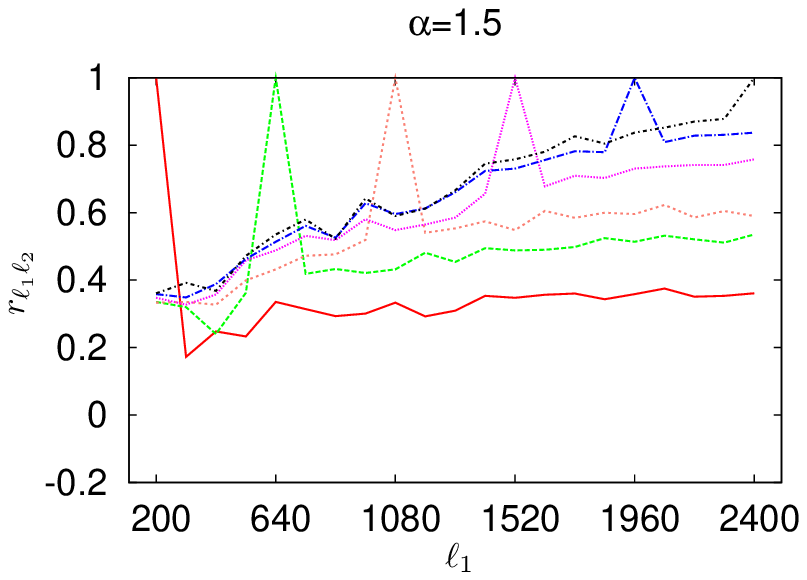} 
 \caption{The correlation coefficient ${\rm r}_{\ell_{1}\ell_{2}}$ of the log-normal covariance matrix as a function of the wave vector $\ell_{1}$ for various values of $\ell_{2}$ for different $\alpha$. The value of $\ell_{2}$ corresponding to each curve can be read off from the point where the curve attains the value $1$.}
 \label{fig:off}
\end{figure}

The validity of the Gaussian approximation furthermore requires the vanishing of the off-diagonal elements of the covariance matrices. We define the correlation coefficient 
\begin{equation}
 {\rm r}_{\ell_{1}\ell_{2}}= C_{\ell_{1}\ell_{2}}/\sqrt{C_{\ell_{1}\ell_{1}} C_{\ell_{2}\ell_{2}}} \;,
\end{equation}
which by definition is unity for the diagonal elements, and plot this in Fig.~\ref{fig:offdia} for different values of $\alpha$. As expected, the off-diagonal elements are compatible with zero for the Gaussian case, and are also very small for the log-normal fields with $\alpha\lesssim 1$. For larger values of $\alpha$, the off-diagonal elements of the correlation coefficient become significantly non-zero, clearly indicating the breakdown of the Gaussian approximation. The amplitude of the off-diagonal elements increases with $\ell_1$ and $\ell_2$, i.e., for smaller scales where the non-Gaussianity of the fields is stronger. To show this behaviour in more detail, Fig.~\ref{fig:off} shows the correlation coefficient ${\rm r}_{\ell_{1}\ell_{2}}$ as a function of the wave number $\ell_{1}$ for various values of $\ell_{2}$. For $\alpha=1.0$ the off-diagonal elements are still almost zero, but increase for $\alpha>1.0$ with large $\ell_{1}$ and $\ell_{2}$. 

\begin{figure}[t]
 \begin{minipage}[h]{.49\linewidth}
  \includegraphics[width=\linewidth]{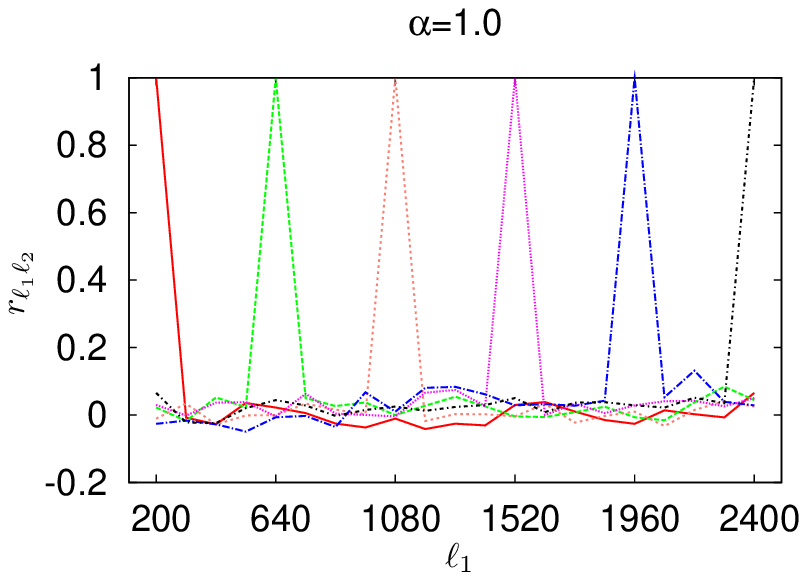}
 \end{minipage}
 \begin{minipage}[h]{.49\linewidth}
  \includegraphics[width=\linewidth]{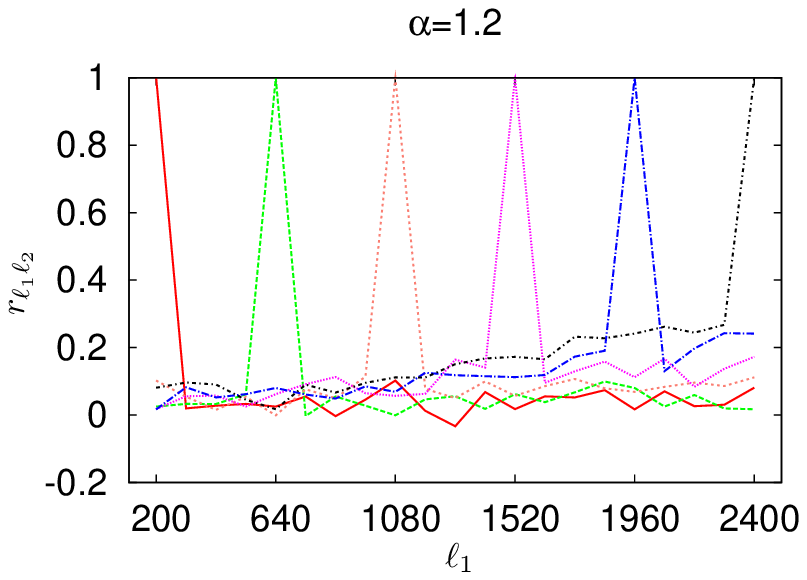}
 \end{minipage}
\end{figure}

Combining the results from studying the diagonal and non-diagonal elements of the bispectrum covariance for equilateral triangles, we conclude that the Gaussian approximation provides, {on the scales considered (down to eight times the pixel scale), an accurate description of the bispectrum covariance for degrees of non-Gaussianity corresponding to $\alpha\lesssim 0.5$, and a very reasonable approximation even for $\alpha\lesssim 1$. For larger degrees of non-Gaussianity, the Gaussian approximation starts to break down, and higher-order terms become important for calculating the bispectrum covariance. In general, the Gaussian approximation works better for large scales than for small scales, as non-Gaussianity contributes more on small scales and therefore the Gaussian approximation should be worse for this case.

\subsubsection{Degree of non-Gaussianity for cosmic shear fields}
\label{subsubsec:Degree of non-Gaussianity of the Universe}
In order to relate our results to future studies of third-order statistics of cosmic shear fields, we need to obtain an estimate of the degree of non-Gaussianity, here parametrised in terms of $\alpha$, of realistic shear (or convergence) fields. As shown by \citet{2002ApJ...571..638T} and \citet{2011arXiv1105.3980H}, the cosmic shear convergence distribution is rather well described by a log-normal; from their results we can estimate characteristic values of $\alpha$. 

A comparison of the probability distributions used in \citet{2002ApJ...571..638T} and in \citet{2011arXiv1105.3980H} with the one used for our investigations (see Eq.~(\ref{eq:pdln})) shows that the probability distributions are equivalent, except that we have fixed $A$ to get unit variance for the log-normal field. The convergence field $\kappa$ of \citet{2002ApJ...571..638T} and \citet{2011arXiv1105.3980H} corresponds to the log-normal field $L$ used in this paper. Furthermore, the parameter $|\kappa_{\rm min}|$ or $\kappa_{0}$ corresponds to the parameter $A$, which was introduced in Sect.~\ref{subsec:Log-normal random fields}. Finally, their quantity $\sigma$ is equivalent to the non-Gaussianity parameter
$\alpha$. Thus, the degree of non-Gaussianity $\alpha$ of the convergence field $\kappa$ can directly be estimated from the $\sigma$ computed by \citet{2002ApJ...571..638T} and
\citet{2011arXiv1105.3980H}.
 
For a source redshift of $z=1$, a smoothing angle of $\theta=4\farcm0$ and a LCDM-model one can compute from \citet{2002ApJ...571..638T} that the degree of non-Gaussianity of the convergence field $\kappa$ corresponds to $\alpha\approx 0.437$. The result we obtain from \citet{2011arXiv1105.3980H} is $\alpha\approx 0.658$ for unsmoothed fields. Here the smoothing is implicitly determined by the parameters in the cosmological simulation, roughly to $\sim3^{\prime\prime}\!\!.5$. Evidently, the value of $\alpha$ is not a strong function of smoothing scale here. In either case, the non-Gaussianity of cosmic shear convergence fields are estimated to correspond to $\alpha<1$, from which we conclude that the Gaussian approximation for the covariance bispectrum is likely to be applicable in cosmic shear studies, at least down to angular scales of $\sim30^{\prime\prime}$.

\subsection{Non-equilateral triangles}
\label{subsec:Non-equilateral triangles}
The results shown so far are based on calculations of the bispectrum for equilateral triangles, to have just a single parameter. However, to get a more complete view on the range of applicability of the Gaussian approximation, the bispectrum must in principle be calculated for all possible triangle shapes. Here, we will consider two more shapes of triangles; for the first case, which we term `truncated triangles', we set $\ell_1=\ell_2=\ell$, $\ell_3=\ell/3$, whereas in the second case, denoted `sharp triangles', we set
$\ell_1=\ell_2=\ell$, $\ell_3=1.5 \ell$. For both cases, we calculate the covariance of the bispectrum, as was done in the equilateral case before.
\begin{figure}[t]
 \includegraphics[width=\linewidth]{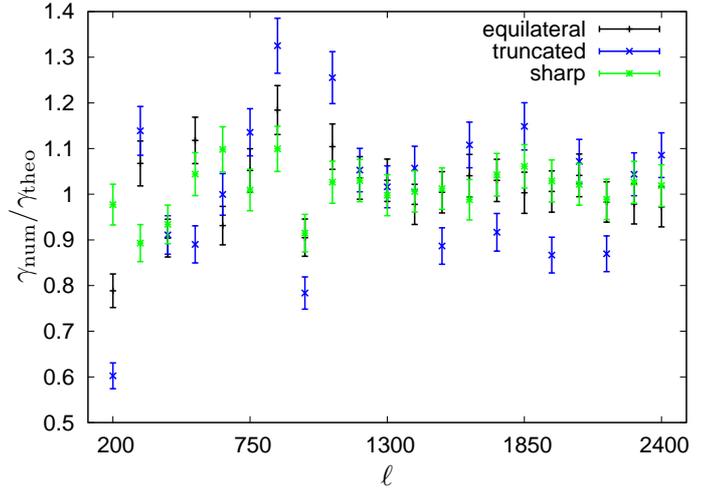} 
 \caption{The relation between the theoretically calculated geometrical factor and the numerically computed geometrical factor $\gamma_{\rm num} / \gamma_{\rm theo}$ for one Gaussian random field (with $1000$ realisations) for different triangle shapes, all of which have $\ell_1=\ell_2=\ell$: equilateral ($\ell_3=\ell$), truncated ($\ell_3=\ell/3$) and sharp ($\ell_3=1.5\ell$) triangles. Error bars are computed by Gaussian error propagation.}
 \label{fig:gammarelsharp}
\end{figure}

In Fig.~\ref{fig:gammarelsharp}, we first check the accuracy of the analytical estimate for the Gaussian case, i.e., the ratio $\gamma_{\rm num}/\gamma_{\rm theo}$ is shown for equilateral triangles (black), truncated triangles (blue) and sharp triangles (green). As seen before, the Gaussian approximation provides a good description of the bispectrum covariance in the case of Gaussian fields. There is a larger scatter around $1$ for truncated triangles, and a smaller scatter for sharp triangles, than for equilateral triangles. This is due to the fact that truncated triangles probe the field properties on larger angular scales than equilateral triangles; the opposite is true for sharp triangles. Accordingly there are fewer (more) triangles to average over for truncated (sharp) triangles than for equilateral ones.

For a log-normal field with $\alpha=0.5$, we plot the aberration parameter $\varsigma$ in Fig.~\ref{fig:relstandsharp} for the three different kinds of triangles. As can be seen, there is no significant difference for truncated and sharp triangles to the results obtained for equilateral triangles. This simple example thus indicates that the Gaussian approximation for the bispectrum covariance also holds for other triangle shapes. We therefore conclude that the Gaussian approximation can be applied for log-normal fields with a degree of non-Gaussianity $\alpha\le 0.5$. In combination with what was said above, this implies that for applications to cosmic shear, the Gaussian approximation for the bispectrum covariance is sufficiently accurate for the range of scales studied.
\begin{figure}[t]
 \includegraphics[width=\linewidth]{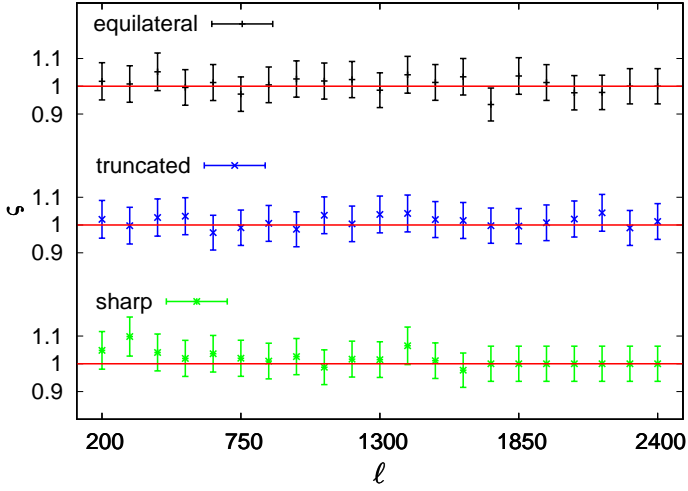} 
 \caption{The aberration parameter $\varsigma(\alpha=0.5)$ for equilateral, truncated and sharp triangles.}
 \label{fig:relstandsharp}
\end{figure}


\section{Summary and Conclusion}
\label{sec:Summary and Conclusion}
In this paper we studied the range of applicability of the Gaussian approximation for the bispectrum covariance, as derived by \citet{2009A&A...508.1193J}, also for non-Gaussian fields. We chose log-normal fields as representative for non-Gaussian fields, also motivated by the fact that the convergence fields underlying cosmic shear are well described by a log-normal one. Our study is based on numerical simulations of Gaussian and log-normal fields with different degrees $\alpha$ of non-Gaussianity. Using 1000 realisations of these
fields, their bispectrum was calculated, and the bispectrum covariance was estimated from the sample variance.

In our simulations, $\alpha$ is a non-Gaussianity parameter of the transformation (\ref{eq:trafo}). This formula is applied to the pixel values in realisations of Gaussian fields on a grid. For convenience, the pixel value variance was chosen to be unity in the Gaussian and in the log-normal case. We could as well have chosen the pixel variance in the Gaussian and non-Gaussian field to be $\sigma^2$ without changing the essential results of this paper because this would have just modified $A$ in Eq. (\ref{eq:logvariance}). This would have rescaled the entire field by a mere constant factor, leaving the skewness (Appendix) untouched.

The following conclusions are drawn for angular scales down to approximately eight times the pixel scale. We have shown that the bispectrum covariance remains almost diagonal, as long as $\alpha\lesssim 1$. Furthermore, the deviation of the diagonal elements from those predicted by the Gaussian approximation are negligible for $\alpha\lesssim 0.5$, and below $10\%$ for $\alpha\lesssim 1$. From simulations of convergence fields in weak lensing shown in \citet{2002ApJ...571..638T} and \citet{2011arXiv1105.3980H}, we found that realistic values of $\alpha$ are not larger than $\sim 0.7$ on angular scales of about $4^{\prime\prime}$; hence, the Gaussian approximation is expected to yield an accurate description of the bispectrum covariance -- which is fortunate, given the complexity of the full expression of this covariance.

Whereas the results depend on the assumed statistical properties of the non-Gaussian field -- here assumed to be log-normal -- as well as on the shape of the power spectrum, which we chose somewhat arbitrarily, we expect that the accuracy of the Gaussian approximation for the bispectrum covariance depends predominantly on the degree of non-Gaussianity (here parametrised by $\alpha$, but we could equivalently use the skewness of the field -- see Appendix).

\begin{acknowledgements}
 We thank Benjamin Joachimi, Jan Hartlap, Xun Shi and Stefan Hilbert for many fruitful discussions and their help in this project. This work was supported by the Deutsche Forschungsgemeinschaft in the framework of the Collaborative Research Center TR33 `The Dark Universe'.
\end{acknowledgements}



\appendix
\section{Skewness of the log-normal field}
Besides the mean and the variance, higher order moments of the log-normal random field $L$ can be calculated easily, for example the third-order moment
\begin{equation}
 \left\langle L(\vec{x})^{3} \right\rangle = A^{3} \left(c^{2}-1\right)^{2}\left(c^{2}+2\right)\,,
\end{equation}
which is of special interest as it is related to the skewness parameter $\Gamma^{(3)}$ by:
\begin{equation}
 \Gamma^{(3)} = \frac{\left\langle L(\vec{x})^{3} \right\rangle}{\left\langle L(\vec{x})^{2} \right\rangle^{3/2}} = \left( c^{2}+2\right) \sqrt{c^{2}-1}\,.
 \label{eq:Gamma3}
\end{equation}
As the skewness depends ultimately only on $\alpha$, the latter can be regarded as to set the skewness of the log-normal field $L(\vec{x})$, which can also be visualised by the Taylor expansion of Eq.~(\ref{eq:Gamma3}),
\begin{equation}
 \Gamma^{(3)} = 3\alpha + \frac{7\alpha^{3}}{4} + \frac{29 \alpha^{5}}{32} + {\cal O}(\alpha^{7})\,.
\end{equation}

\end{document}